\documentclass[a4paper,fleqn]{cas-dc}

\usepackage[sort&compress,numbers]{natbib}
\usepackage{eqnarray,amsmath}
\newcommand{\sizethree}{0.43\textwidth}
\newcommand{\sizetfour}{0.23\textwidth}

\newcommand{\NO}{\textrm{NO}}
\newcommand{\NON}{\NO$_{\textrm{n}}$ }
\newcommand{\AFT}{\textrm{AFT}}
\newcommand{\FER}{\textrm{FER}}
\newcommand{\FERN}{\FER$_{\textrm{n}}$}
\newcommand{\AFTA}{\AFT$_{\textrm{A}}$}
\newcommand{\AFTB}{\AFT$_{\textrm{B}}$}
\newcommand{\AFTC}{\AFT$_{\textrm{C}}$}

\begin{document}
\let\WriteBookmarks\relax
\def\floatpagepagefraction{1}
\def\textpagefraction{.001}

\shorttitle{Magnetic and charge orders on the triangular lattice}
\shortauthors{K. J. Kapcia \& J. Bara\'nski}

\title[mode = title]{Magnetic and charge orders on the triangular lattice: 
Extended Hubbard model  with intersite Ising-like magnetic interactions in the atomic limit}

\author[1]{Konrad Jerzy Kapcia}[%
                    orcid=0000-0001-8842-1886]
\ead{konrad.kapcia@amu.edu.pl}
\credit{Conceptualization, Methodology, Software, Validation, Formal analysis, Investigation, Resources, Data curation, Writing - Original draft preparation, Writing - Review \& Editing, Visualization, Supervision, Project administration, Funding acquisition}

\author[2]{Jan Bara\'nski}[%
                        orcid=0000-0002-0963-497X]
\ead{j.baranski@law.mil.pl}
\credit{Validation, Formal analysis, Investigation, Writing - Original draft preparation, Writing - Review \& Editing, Visualization}

\address[1]{Institute of Spintronics and Quantum Information, Faculty of Physics, Adam Mickiewicz University in Pozna\'n, ul. Uniwersytetu Pozna\'{n}skiego 2, 61614 Pozna\'{n}, Poland}
\address[2]{Department of General Education, Polish Air Force University, ul. Dywizjonu 303 nr 35, 08521 Deblin, Poland} 

\begin{abstract}
In the work, we investigated a generalized model of the fermionic lattice gas in the form of the extended Hubbard model with intersite Ising-like interactions (both antiferromagnetic and ferromagnetic)  at the atomic limit on the triangular lattice.
In the ground state, we find the exact phase diagram as a function of $\mu$.
Within the mean-field decoupling of the intersite term and exact treatment of onsite interaction, we found also the diagrams for $T\geq 0$ including metastable phases.
For antiferromagnetic coupling, we find that nontrival ordered phase can exist with coexistence of charge and metamagnetic ordering.
The transition between the ordered phase and the nonordered phase can be discontinuous as well continuous depending on the model parameters.
Moreover, the ordered phase can coexist with the nonordered phase in phase separated states for fixed electron concentration.
Additionally, the ranges of phases metastability are determined in the neighborhood of the discontinuous transitions.
\begin{itemize}
\item[$\ $]{\large{\textsc{Highligts:}}}
\item The extended Hubbard model with intersite Ising-like interactions is investigated.
\item Orderings on the triangular lattice with geometrical frustration are analyzed. 
\item The exact ground state for fixed chemical potential is determined.
\item Mean-field field solutions of the model for finite temperatures are found and discussed. 
\item Complex phase diagram for both signs of magnetic interactions are determined.
\end{itemize}
\end{abstract}





\maketitle

\section{Introduction}

The strongly correlated system exhibits many intriguing phenomena from band renormalization to very complex diagrams with phases involving charge, spin, orbital or superconducting orders \cite{MicnasRMP1990,GeorgesPMP1996,ImadaRMP1998,FreericksRMP2003}.
Moreover, the ability of controlling the interactions in the ultra-cold fermionic gases on the optical lattices via Feshbach resonances releases the new possibilities  for experimental studies of variety of unconventional systems \cite{GeorgescuRMP2014,DuttaRPP2015}, in particular, in systems with geometrical frustration \cite{StruckScience2011,MishraNJP2016}. 

Inspired by the rich structure of the simple model of charged ordered insulators on the triangular lattice \cite{KapciaJSNM2019,KapciaNano2021,KapciaJMMM2022}, in this work, we investigate a generalized model of the strongly correlated fermionic lattice gas, which can be also considered as  a generalization of the standard $S=1/2$ Ising model  \cite{HoutappelPhys1950a,HoutappelPhys1950b,CampbellPRA1972,KaburagiJJAP1974,MetcalfPLA1974,MihuraPRL1977,KaburagiJPSJ1978}.
The studied Hamiltonian has the following form \cite{KlobusAPPA2010,MurawskiAPPA2012,ManciniOP2012,ManciniEPJB2013,MurawskiAPPA2014,MurawskiAPPA2015,KapciaPhysA2015,Murawski2016}:
\begin{equation}\label{eq:ham}
\hat{H} = 
U \sum_i \hat{n}_{i,\uparrow} \hat{n}_{i,\downarrow} + 
2 J \sum_{\langle i, j \rangle} \hat{s}_i \hat{s}_j - 
\mu \sum_i \hat{n}_i,
\end{equation}
where 
$\hat{n}_{i,\sigma} = \hat{c}_{i,\sigma}^{\dag} \hat{c}_{i,\sigma}^{\ } $, 
$\hat{s}_i = \frac{1}{2} \left( \hat{n}_{i,\uparrow} - \hat{n}_{i,\downarrow} \right)$, and 
$\hat{c}_{i,\sigma}^{\dag}$  ($\hat{c}_{i,\sigma}^{\ } $) denotes the creation (annihilation) operator of a fermion with spin $\sigma$ ($\sigma \in \{ \uparrow, \downarrow \}$) at lattice site $i$.
$U$ is onsite Hubbard interaction, $J$ is intersite Ising-like magnetic interaction between nearest-neighbor sites $i$ and $j$, and $\mu$ denotes the chemical potential.
$\sum_{\langle i, j \rangle}$ indicates summation over nearest-neighbor sites independently.
Note that, for $U\rightarrow + \infty$ and $n=1$ (or $\mu=U/2$; the half-filling), model (\ref{eq:ham}) reduces to the Ising model (with two possible states $s_i= \pm 1/2$ on a single lattice site) in the absence of the external field (in a general case, in the external field of $\mu-U/2$).

The model has been intensively analyzed on the alternate (bipartite) lattices.
The case of one-dimensional chain without and with the external magnetic field was investigated in \cite{ManciniOP2012,KapciaAPPA2015a} and \cite{ManciniEPJB2013}, respectively.
Classical Monte Carlo simulations were performed for the model on two-dimensional square lattice system \cite{MurawskiAPPA2012,MurawskiAPPA2014,MurawskiAPPA2015,KapciaPhysA2015}.
Finally, rigorous results (in the limit of $d\rightarrow + \infty$) for model (\ref{eq:ham}) considered on the hypercubic lattices were obtained in \cite{RobaszkiewiczPSSB1975,RobaszkiewiczAPPA1979,KlobusAPPA2010,KapciaPhysA2015} with the use of the variational approach with the mean-field approximation for the intersite term and a rigorous treatment of the local interactions. 

The investigated model exhibits the particle-hole symmetry, thus, it is enough to analyze the model for $\mu<U/2$ (or $n<1$).
Note that for the model on the hypercubic lattices, the $J \leftrightarrow -J $ symmetry  occurs \cite{KapciaPhysA2015}, but for the triangular lattice it is no longer valid.

\section{Mean-field expressions for finite temperatures}

The method based on the mean-field approximation for the intersite terms (where as the onsite terms treated exactly) is applied for model (\ref{eq:ham}) to find thermodynamic behavior of the model at arbitrary temperatures (details of the method can be find in, e.g.,  \cite{KapciaPhysA2015,KapciaNano2021,MicnasPRB1984}).
The grand canonical potential is determined as 
\begin{equation}
\label{eq:omegatemp}
\omega = -\frac{1}{3} \sum_{\alpha} \psi_\alpha s_\alpha - \frac{1}{3\beta} \sum_{\alpha} \ln Z_{\alpha},
\end{equation}
where $\beta = 1/(k_{B} T)$ is inverted temperature ($k_B$ is the Boltzman constant), coefficients $\psi_{\alpha}$ for $\alpha \in \{ A, B, C\} $ are defined as $\psi_A = J (s_B+s_C)$, $\psi_B = J (s_A+s_C)$, $\psi_C = J (s_A+s_B)$, and
\begin{equation}
\label{eq:zalpha}
Z_{\alpha} = 1 + \exp \left[ \beta \left( 2\mu - U \right) \right]  + 2 \exp \left( \beta \mu \right) \cosh \left( \beta \psi_{\alpha} \right).
\end{equation}
Equations for concentrations $n_{\alpha} = \langle \hat{n}_i \rangle_{i \in \alpha}$ and magnetizations $s_{\alpha} = \langle \hat{s}_i \rangle_{i \in \alpha}$ in each sublattice $\alpha\in \{A, B, C \}$ are obtained as
\begin{eqnarray}
\label{eq.concetrations}
n_{\alpha} & = & \frac{2}{Z_{\alpha}} 
\left\{ \exp \left[ \beta \left(2\mu-U \right) \right] + \exp \left( \beta \mu \right) \cosh \left( \beta \psi_{\alpha} \right) \right\},\qquad \\
\label{eq.magnetizations}
s_{\alpha} & = & \frac{1}{Z_{\alpha}} \exp \left( \beta \mu \right) \sinh \left( - \beta \psi_{\alpha} \right).
\end{eqnarray}
One can also derived expressions for other quantities, e.g., for double occupancy $D_{\alpha} =  \langle \hat{n}_{i,\uparrow} \hat{n}_{i,\downarrow} \rangle_{i \in \alpha}$ and for local magnetic moment $\gamma_{\alpha} =  \langle | \hat{s}_{i} | \rangle_{i \in \alpha}$  as
\begin{equation}\nonumber
D_{\alpha} = \frac{1}{Z_{\alpha}}\exp{(2\mu-U)}, \  \gamma_{\alpha} =  \frac{1}{Z_{\alpha}} \exp \left( \beta \mu \right) \cosh \left( \beta \psi_{\alpha} \right).
\end{equation}
These quantities are related by $D_{\alpha}+\gamma_{\alpha} = n_{\alpha}/2$.
They also determine the fraction of single and double occupied sites, respectively, in each sublattice.
$2D_{\alpha}$ and $2\gamma_{\alpha}$ stand for the numbers of locally paired and unpaired particles, respectively, in each sublattice. 
The total particle concentration $n$ is defined as $n=\sum_{\alpha}n_\alpha /3$.

Note that the right hand side of Eq. (\ref{eq.concetrations}) [as well as Eq. (\ref{eq:zalpha})] does not include $n_{\alpha}$'s. 
Thus, effectively one needs to solve self-consistently a set of only three non-linear equations (\ref{eq.magnetizations}) to determine $s_A$, $s_B$, and $s_C$ (for fixed model parameters $\mathcal{P}=\{\mu, U, J, T\}$). 
$n_{\alpha}$'s are determined by $s_{\alpha}$'s and parameters $\mathcal{P}$.
However, the set can have several solutions for $s_{\alpha}$ and one needs to find the solution corresponding to the lowest $\omega$ determined by (\ref{eq:omegatemp}). 
Due to the symmetry of the system, there are several solutions, which are equivalent with $(s_A,s_B,s_C)$ solution, e.g., $(-s_A,-s_B,-s_C)$, $(s_B,s_C,s_A)$, $(s_C,s_A,s_B)$, $(s_B,s_A,s_C)$.
In total, there can be $12$ equivalent solutions, but their number is reduced if there are some particular relations between $s_\alpha$'s, e.g., if $s_A=-s_B$.

From numerical analysis of equations (\ref{eq.magnetizations}) for antiferromagnetic interactions ($J>0$) presented in Sec. \ref{section:fintempaf}, one gets that the solutions, which correspond to the lowest $\omega$, have the form of $(s,-s,0)$ (there is $6$ such equivalent solutions).
If $|s|>0$, we call this phase the {\AFT} phase (antiferromagnetic alignment of average magnetic moments in two sublattices, in the third one - no magnetization), whereas, if $s=0$, it is the normal (non-ordered, {\NO}) phase. 
Note that,  if, e.g., $s_A = -s_B$ for some $\mathcal{P}$, one gets immediately  that $\psi_{C} = 0$ and, from (\ref{eq.magnetizations}), $s_C=0$ for that $\mathcal{P}$.
In such case, Eqs. (\ref{eq.magnetizations}) and (\ref{eq:omegatemp}) take the forms
\begin{equation}\label{mag:AFT}
s = \frac{\sinh (\beta J s)}{2 \left[ \cosh(\beta \bar{\mu}) \exp(-\beta U/2) + \cosh (\beta J s) \right]}, 
\end{equation}
\begin{eqnarray}
\omega & =   - \bar{\mu} -\frac{2}{3} J s - \frac{1}{3 \beta} \ln \left[ 2 \cosh (\beta \bar{\mu}) + 2 \exp (\beta U/2)\right] \nonumber  \\
 & -   \frac{2}{3\beta} \ln \left[   2 \cosh (\beta \bar{\mu}) + 2 \exp (\beta U/2) \cosh (\beta J s )\right]. \qquad
\end{eqnarray}

In the case of the ferromagnetic coupling ($J<0$), the lowest $\omega$ solutions are found as $(s,s,s)$ (two equivalent solutions), and the only ordered phase occurring is the ferromagnetic ({\FER}) phase (for $|s|>0$).
In such a case, equations (\ref{eq.concetrations})--(\ref{eq.magnetizations}) reduce to those obtained in Ref.~\cite{KapciaPhysA2015}. 
Then, Eqs. (\ref{eq.magnetizations}) and (\ref{eq:omegatemp}) take the forms
\begin{equation}\label{mag:FER}
s = \frac{\sinh (- \beta 2 J s)}{2 \left[ \cosh(\beta \bar{\mu}) \exp(-\beta U/2) + \cosh (\beta 2 J s) \right]}, 
\end{equation}
\begin{align}
\omega & =   - \bar{\mu} -2Js   \\
 & -  \frac{1}{\beta} \ln \left[ 2 \cosh (\beta \bar{\mu}) + 2 \exp(\beta U /2) \cosh (\beta 2J s) \right]. \nonumber
\end{align}
Thus, for the ferromagnetic coupling $J$ within the mean-field approximation, the results for model (\ref{eq:ham}) on the triangular lattice and on the bipartite lattices are the same (cf. also Refs. \cite{KlobusAPPA2010,MurawskiAPPA2012}).
Note that, for the model on the alternate lattices (in the absence of the external magnetic field, analyzed in detail in Refs. \cite{KlobusAPPA2010,MurawskiAPPA2012,KapciaPhysA2015}), the sign of the nearest-neighbor magnetic coupling is irrelevant for the phase boundaries and only type of order in the ordered phase is different: if $J<0$ the ferromagnetic order exists, whereas for $J>0$ the antiferromagnetic order appears.

One can notice that Eqs. (\ref{mag:AFT})  and (\ref{mag:FER}) (variables with {\AFT} and {\FER} labels, respectively) give the same solution for $s$ ($s_{\FER} = s_{\AFT}$) if $J_{\AFT} = - J_{FER} >0 $ and 
\begin{equation}
\label{eq:transFERAFT1}
T_{\AFT} = \frac{T_{\FER}}{2}, \ \bar{\mu}_{\AFT} = \frac{\bar{\mu}_{\FER}}{2}, \ U_{\AFT} = \frac{U_{\FER}}{2}
\end{equation}
and $\beta_{\AFT} = 2\beta_{\FER}$.
The relation between concentration in both ordered phases can be obtained from Eq. (\ref{eq.concetrations}) with using (\ref{eq:transFERAFT1}) and one gets 
\begin{eqnarray}\label{eq:transFERAFT2}
2 n_{\FER} \left( T_{\FER}, U_{\FER}, \bar{\mu}_{\FER}  \right)= \qquad \qquad \\
 3n_{\AFT} \left( T_{\AFT}, U_{\AFT}, \bar{\mu}_{\AFT}  \right) - n_{\NO} \nonumber,
\end{eqnarray}
where $n_{\NO}$ is the total  concentration in the non-ordered phase and $n_{\NO} \left( T_{\FER}, U_{\FER}, \bar{\mu}_{\FER}  \right) = n_{\NO} \left( T_{\AFT}, U_{\AFT}, \bar{\mu}_{\AFT}  \right) $.
Note also  that conditions for the order-disorder boundaries, i.e.,  $\omega_{\NO} = \omega_{\AFT}$ and $\omega_{NO} = \omega_{\FER}$ (where $\omega_{\NO}$ denotes $\omega$ for $s=0$) also can be transformed into each other using (\ref{eq:transFERAFT1}).
Because of that, in the following, we will focus mainly on model (\ref{eq:ham}) with antiferromagnetic interactions (i.e., $J>0$).

Note that both Eqs. (\ref{mag:AFT}) and (\ref{mag:FER}) can be rewritten as 
\begin{equation}\label{mag:both}
s = \frac{\sinh ( \beta k J s)}{2 \left[ \cosh(\beta \bar{\mu}) \exp(-\beta U/2) + \cosh (\beta k J s) \right]}, 
\end{equation}
where (i) $k=1$ if $J>0$ (the {\AFT} phase) or (ii) $k=-2$ if $J<0$ (the {\FER} phase).
Using similar procedure as in \cite{RobaszkiewiczAPPA1979,MicnasPRB1984,KapciaNano2021,KapciaJMMM2022}, taking the limit of $s\rightarrow 0$ of both sides of the equation above (divided by $s$ before) and applying the de l’Hospital theorem) one gets the expression for the second order temperature $T_{c}$ for order-disorder transition as
\begin{equation}\label{eq:secondorder}
k_{B} T_{c}/J = (k t) \left(1+ 2t+ t^2a\right)^{(-1)},
\end{equation}
where $t =\exp \left( \beta_c \mu \right)$, $a = \exp \left(- \beta_c U \right)$, $\beta_c = 1/(k_B T_c)$. 
At the boundary the relation between $\mu$ and $n$ is expressed by $\mu = (1/\beta_{c}) \ln (x)$, where $x = \left(n-1+ \sqrt{y}\right)/\left((2-n)a \right)$ and $y=(1-n)^2 + n(2-n) a$.
All found continuous  boundaries on the diagrams presented in Sec. \ref{section:fintempaf}  fulfill condition (\ref{eq:secondorder}).
Moreover, the solutions of (\ref{eq:secondorder}) coincide with the metastability boundaries for the NO phase.

\begin{figure}
    \centering
    \includegraphics[width=\sizetfour]{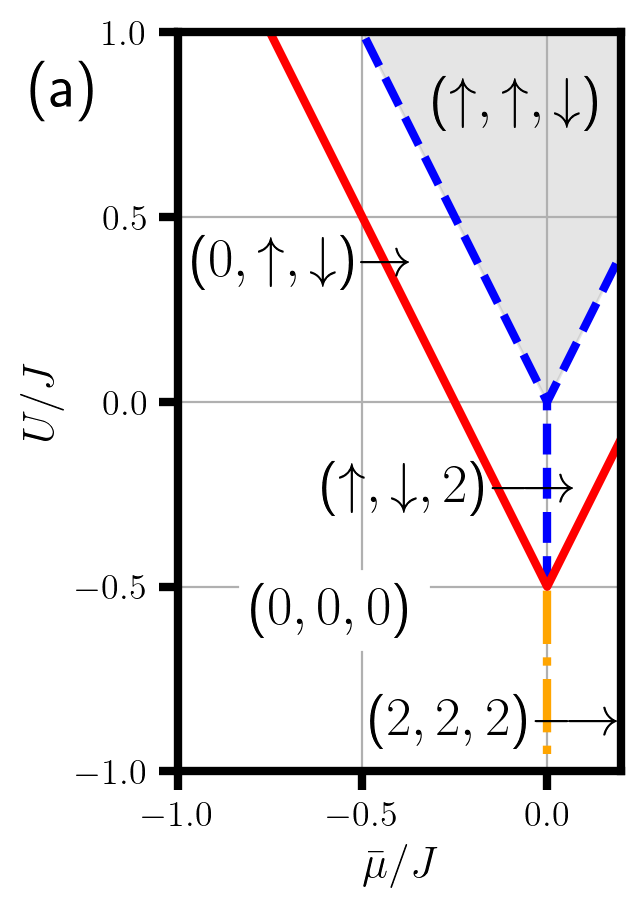}
    \includegraphics[width=\sizetfour]{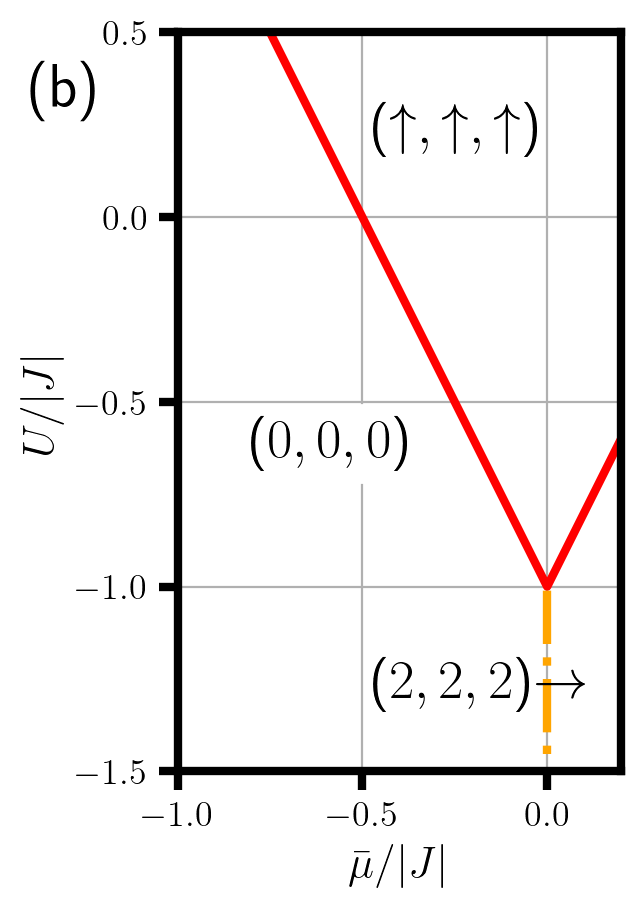}
    \caption{\label{fig:GSmi}%
 	Ground state phase diagram of the model for (a) $J>0$ and (b) $J<0$ (as a function of $\bar{\mu}= \mu - U/2$, rigorous results).
 	The regions are labeled by configuration in the elementary block (defined in text).
        Different lines denote the boundaries with infinite (macroscopic) degeneration [dashed (blue) line] and finite 
        degeneration [solid (red) and and dot-dashed (orange) lines]. 
        Dot-dashed (orange) lines denotes the boundaries which vanishes at any infinitesemelly small $T>0$ (within the MFA).
        Grey shadow denotes a region with infinite degeneration. 
        The regions are labeled by the arrangement in the elementary block.}
\end{figure}

\section{The ground state results}

To determine the ground state phase diagram as a function of the chemical potential the elementary block method is used, which is described in detail in Refs. \cite{JedrzejewskiZFPB1982,JedrzejewskiZFPB1985,BrandtZPB1986,JEDRZEJEWSKIPhysA1994,BorgsJPA1996,KapciaNano2021,KapciaPRE2017}. 
In these works, it was used for the atomic limit of the extended Hubbard model with intersite density-density interactions, but it can be  also applied to model (\ref{eq:ham}) to determine the grand canonical potential $\omega_0$ (per site) at $T=0$.  
In the model, at each site the following four states are possible: $0$ (empty), $\uparrow$ (particle with spin-up), $\downarrow$ (particle with spin-down), $2$ (double occupied), which characterize the elementary block. 
The total number of them is $3^4 = 81$, but some of them are equivalent due to spin-inversion symmetry of the model and permutation of the sublattice labels (e.g., $(0,\uparrow, \downarrow)$ is equivalent with $(\downarrow ,\uparrow, 0 )$).

\begin{figure}
    \centering
    \includegraphics[width=\sizetfour]{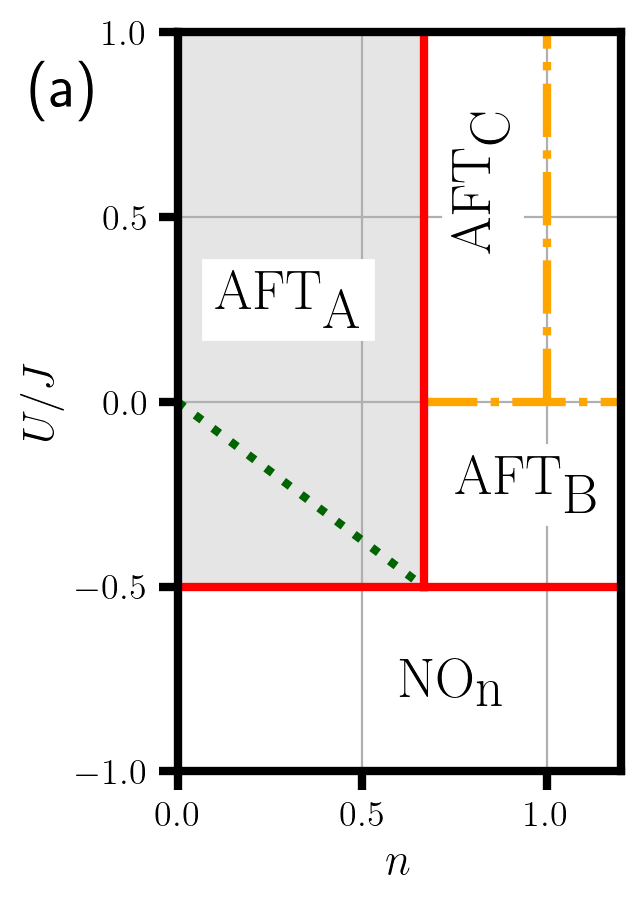}
    \includegraphics[width=\sizetfour]{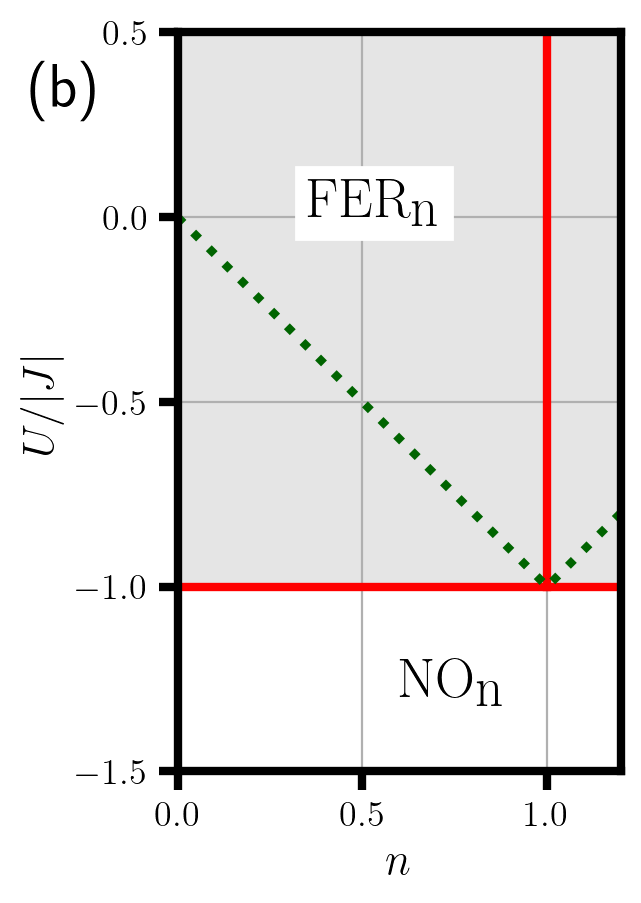}
    \caption{\label{fig:GSn}%
 	Ground state phase diagram of the model for (a) $J>0$ and (b) $J<0$ (as a function of $n$).
        The regions are labeled by homogeneous phases (defined in text).
        Shadows denote regions, where phase separated states have lower energy (at $T\geq 0$).
        The dotted line denotes the boundary between homogeneous phases in the region of the PS state occurrence.
 	Dot-dashed (orange) lines denotes the boundaries, which vanishes at any infinitesimally small $T>0$.}
\end{figure}

The resulting ground state phase diagrams of the model (\ref{eq:ham}) for both signs of $J$ interaction are presented in Fig. \ref{fig:GSmi}.
The following arrangements appear in Fig. \ref{fig:GSmi}(a) (for $J>0$): 
$(0,0,0)$ with $\omega_0=0$; 
$(0,\uparrow,\downarrow)$ with $\omega_0=-J/6 -2\mu/3$; 
$(\uparrow,\uparrow,\downarrow)$ with $\omega_0= -J/6-\mu$;  
$(2,\uparrow,\downarrow)$ with $\omega_0=U/3-J/6 -4\mu/3$; and 
$(2,2,2)$ with $\omega_0=U- 2\mu$.
Note that block $(\uparrow,\uparrow,\downarrow)$ appearing in  Fig. \ref{fig:GSmi}(a) is equivalent to $(\downarrow,\uparrow,\downarrow)$ and $(\uparrow,\downarrow,\downarrow)$. 
Therefore, inside its region, at least two of those three equivalent configurations can be mixed freely. 
Consequently, in the shadowed region, the ground state is infinitely degenerated (and the entropy per site is nonzero).
Similarly, at the dashed boundaries, blocks from neighboring regions can mix with each other freely, e.g., $(0,\uparrow,\downarrow)$ with $(\uparrow,\uparrow,\downarrow)$ (coexistence at the microscopic level).
In the mean-field approximation, in this region, a phase with $s_C=0$ occurs.
For $J<0$ (Fig. \ref{fig:GSmi}(b)), apart from  $(0,0,0)$, and $(2,2,2)$, the region of  $(\uparrow,\uparrow,\uparrow)$ with $\omega_0=J/2-\mu$ is present. 
This is in agreement with \cite{KlobusAPPA2010,KapciaPhysA2015}.

To find the phase diagram at $T=0$ as a function of total particle concentration $n=(n_A+n_B+n_C)/3$, we determine the ground state free energies $f_0=\omega_0 + \mu n$ for the homogeneous phases in the mean-field approximation (cf. also \cite{KapciaPRE2017,KapciaNano2021}).
We use the following denotations for phase properties: $\bar{N} \equiv (n_A,n_B,n_C)$ (describing charge ordering), $\bar{S}\equiv (s_A,s_B,s_C)$ (describing spin ordering), $\bar{D}\equiv (D_A,D_B,D_C)$ (double occupancies).

One gets the following results for appearing homogeneous phases:
(i) nonordered {\NON} with $\bar{N}=(n,n,n)$, $\bar{S}=(0,0,0)$, $\bar{D}=(n/2,n/2,n/2)$ and $f_0 = Un/2$;
three  {\AFT} phases (distinguishable only at $T=0$): 
(ii)  {\AFTA} with $\bar{N}=(0,3n/2,3n/2)$, $\bar{S}=(0,3n/4,-3n/4)$, $\bar{D}=(0,0,0)$ and $f_0 = -3Jn^2/8$;  
(iii) {\AFTB} with $\bar{N}=(3n-2,1,1)$, $\bar{S}=(0,1/2,-1/2)$, $\bar{D}=((3n-2)/2,0,0)$ and $f_0 = U(3n-2)/6-J/6$;
(iv) {\AFTC} with $\bar{N}=(3n-2,1,1)$, $\bar{S}=(0,1/2,-1/2)$, $\bar{D}=(0,0,0)$ and $f_0 = -J/6$;
and (v) ferromagnetic {\FERN} with $\bar{N}=(n,n,n)$, $\bar{S}=(n/2,n/2,n/2)$, $\bar{D}=(0,0,0)$ and $f_0 = Jn^2/2$.
Apart from the {\AFTA} and the {\FERN} phases, all of them are degenerated with phase separates states in which two domains of commensurate phases (with $n=0,2/3,4/3,2$) coexist (mentioned previously for fixed $\mu$ and existing on the boundaries of regions from Fig.  \ref{fig:GSn}).
This degeneracy is removed at finite temperatures.
Note that {\AFTA} and the {\FERN} are unstable in their ranges of occurrence, i.e., $\partial^2 f/ \partial n^2 <0$.
It turns out that, for $J>0$, the PS state (with domains $(0,\uparrow,\downarrow)$ and $(0,0,0)$) exists (with $f_0 = -Jn/4$), whereas for $J<0$, the PS state (with domains $(\uparrow,\uparrow,\uparrow)$ and $(0,0,0)$) has the lowest free energy ($f_0 = Jn/2$) in the areas depicted by shadow in Fig. \ref{fig:GSn}
The dotted line denotes the boundary between homogeneous phases inside the regions of the PS state occurrence.

\section{Finite temperatures phase diagrams ($J>0$)}
\label{section:fintempaf}

\begin{figure}
    \centering
    \includegraphics[width=\sizethree]{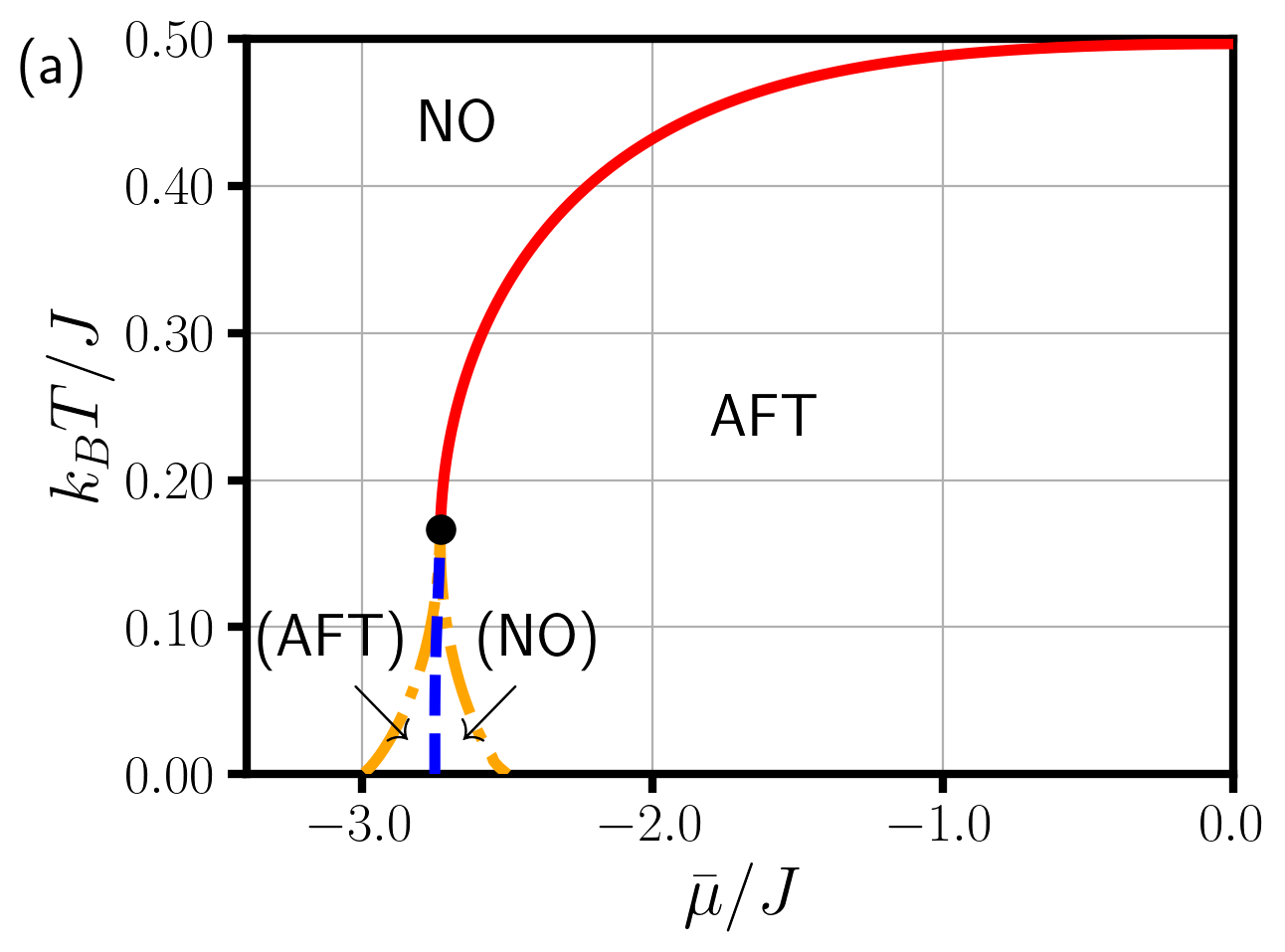}
    \includegraphics[width=\sizethree]{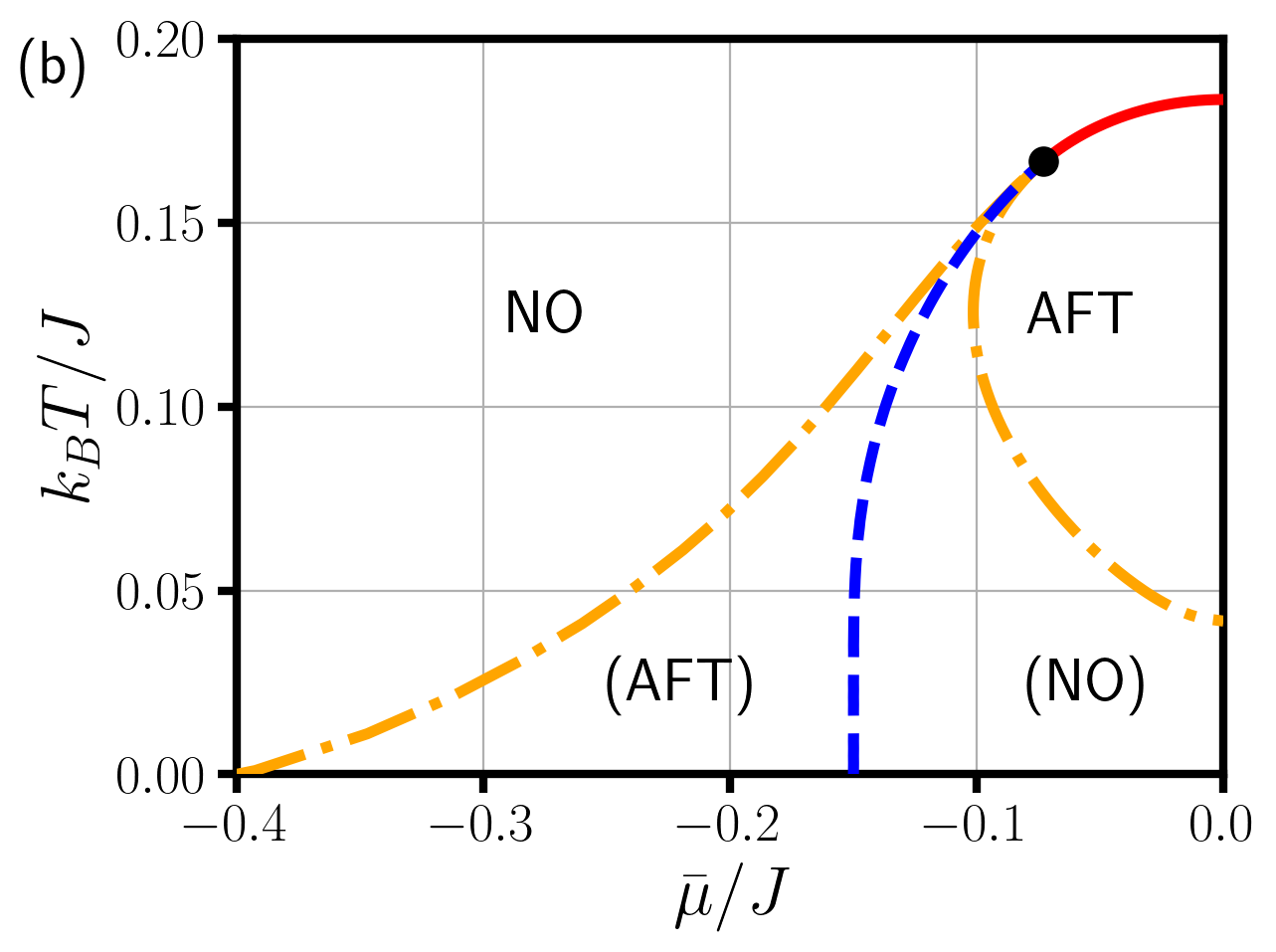}
    \includegraphics[width=\sizethree]{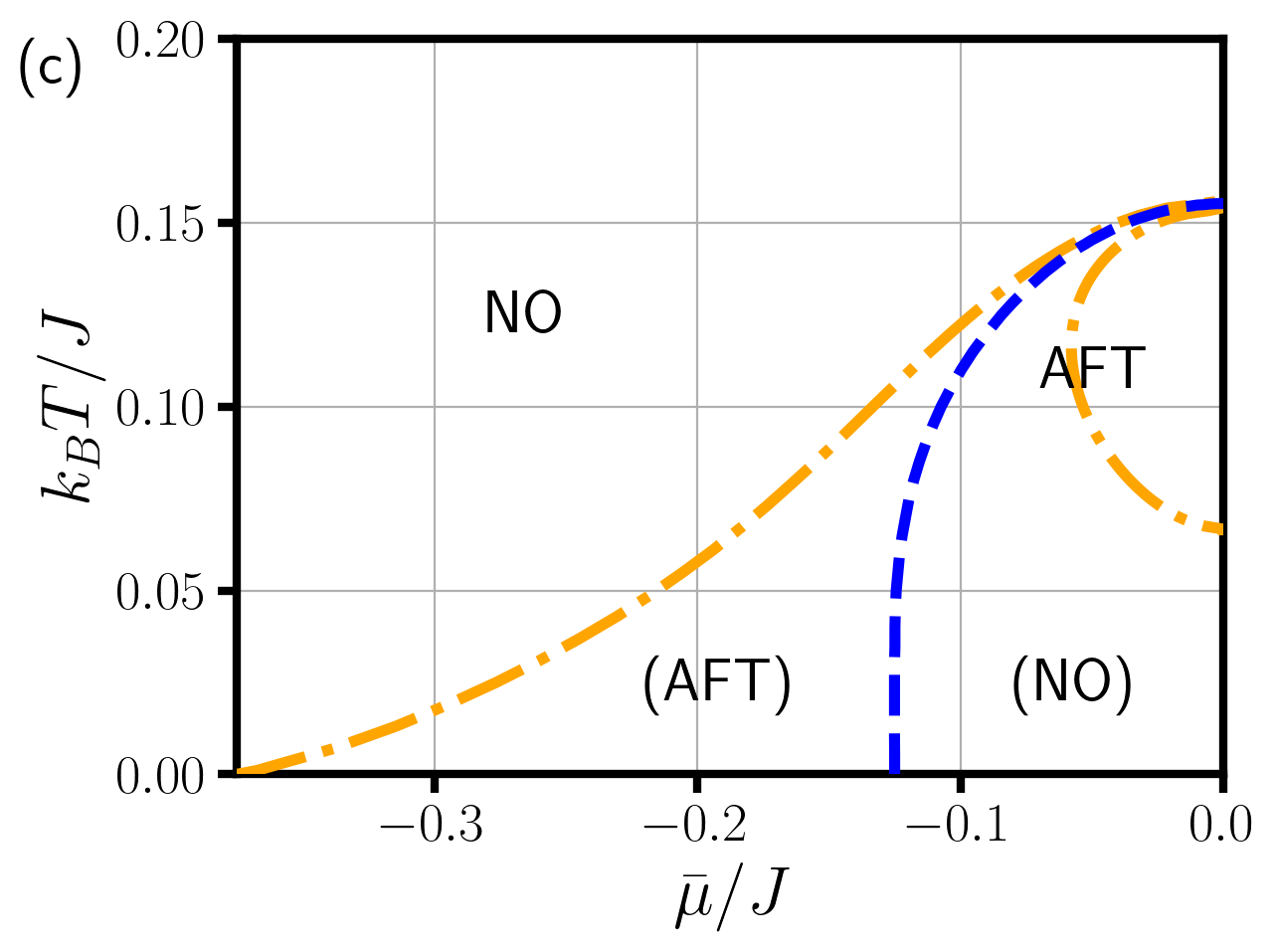}
    \caption{\label{fig:finkTvsmi}%
        Phase diagrams for $U/J=5.00$, $U/J=-0.20$, $U/J=-0.25$ as a function of $\bar{\mu}=\mu-U/2$.
        Solid and dashed lines denote continuous and discontinuous transitions between the {\AFT} and {\NO} phases, whereas dash-dotted lines indicate the boundaries of metastable phase occurrence (names of metastable phases in brackets).
    }
\end{figure}

This section is devoted for a discussion of the phase diagrams of the model, including the metastable phases for $J>0$.
Exemplary phase diagrams as a function of chemical potential are presented in Fig. \ref{fig:finkTvsmi} (cf. also \cite{KapciaPhysA2015}).
In finite temperatures, only two phases occur on the phase diagram: the NO phase and the {\AFT} phase. 
For fixed $\bar{\mu}/J$ and $U/J>0$,  at temperatures above the threshold of $J/(6k_B)$, the transition between the  {\NO} and {\AFT} phases is second order (cf. solid lines in Fig. \ref{fig:finkTvsmi}(a)) and metastable phases does not appear.
The transition changes its order at the tricritical point (TCP) and
for temperatures below $J/(6k_B)$, the transition changes its order to the first one (dashed lines). 
Discontinuity of the transition allows for appearance of metastable phases, i.e., solutions with higher $\omega$'s than those for stable solution with the lowest $\omega$ (both stable and metastable solutions need to be local minima of $\omega(s)$). 
Indeed, we find the regions, in which such phases emerge. 
These regions (in Fig. \ref{fig:finkTvsmi}) are limited by dashed lines from one side and dash-dotted lines from the other. 
The transition temperature is always decreasing function of $|\bar{\mu}|/J$ with its maximum at $\bar{\mu}=0$.

Upon decreasing $U/J$ ratio, the TCP shifts towards $\bar{\mu}=0$. 
Along with this effect, the boundary of metastable NO phases expands towards $\bar{\mu}=0$ (inside the region of the {\AFT} phase stability). 
For a critical value of $U=0$, the metastable {\NO} phase boundary reaches $\bar{\mu}=0$ at zero temperature (result not presented) and the {\NO} phase is metastable for any $\bar{\mu}$ at $T=0$. 
The decrease of $U/J$ below zero causes that the metastable {\NO} phase occurs at $\bar{\mu}=0$ (in the region of the {\AFT} phase stability) even for non-vanishing temperatures (Fig.~\ref{fig:finkTvsmi}(b)). 
At $U/J=-1/3 \ln 2 \approx -0.23$, the second order boundary vanishes and the {\AFT}-{\NO} transition is discontinuous for all range of $\bar{\mu}$ and at any temperature (Fig.~\ref{fig:finkTvsmi}(c)).
Comparing the right sides of Figs.~\ref{fig:finkTvsmi}(b) and \ref{fig:finkTvsmi}(c) one can note that with further decreasing of $U/J$ ratio, the {\AFT} phase region, where metastable {\NO} phase is not present, shrinks (cf. the regions restricted by the right dash-dotted line). 
Below $U/J=-0.279$ this region vanishes completely and in the whole region of the {\AFT} phase stability, the metastable {\NO} phase occurs as well 
(similar effect has been predicted previously for the metastable {\NO} phase inside superconducting phase \cite{KapciaJSNM2014}).
For $U/J< - 1/2$ the {\AFT} region vanishes and only the NO phase is stable. 
The ordered {\AFT} phase does not have the lowest energy for any model parameters (is not stable), but up to $U/J=-1$, the {\AFT} is still metastable near the half-filing ($\bar{\mu}=0$).

\begin{figure*}
    \centering
    \includegraphics[width=\sizethree]{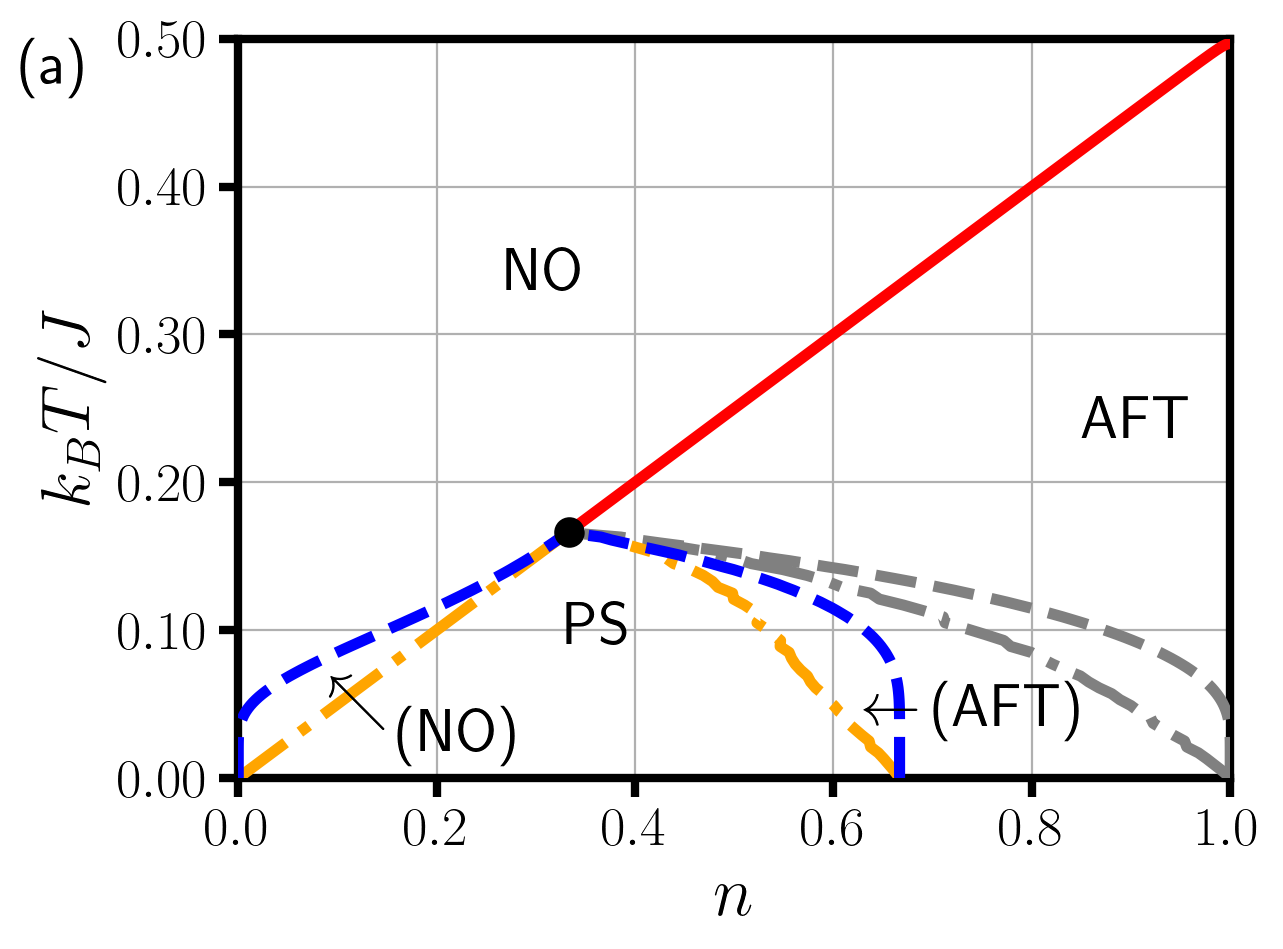}
    \includegraphics[width=\sizethree]{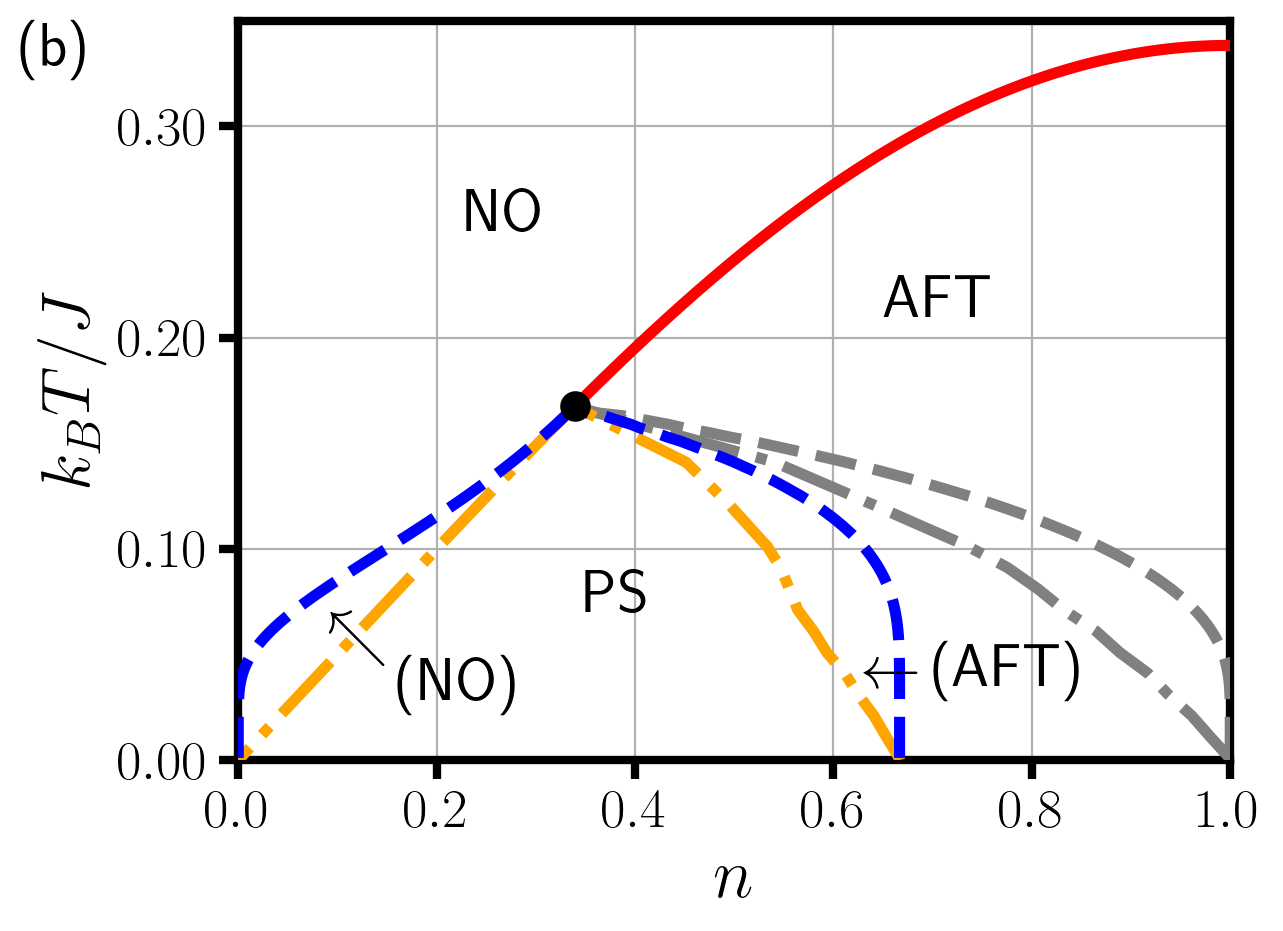}
    \includegraphics[width=\sizethree]{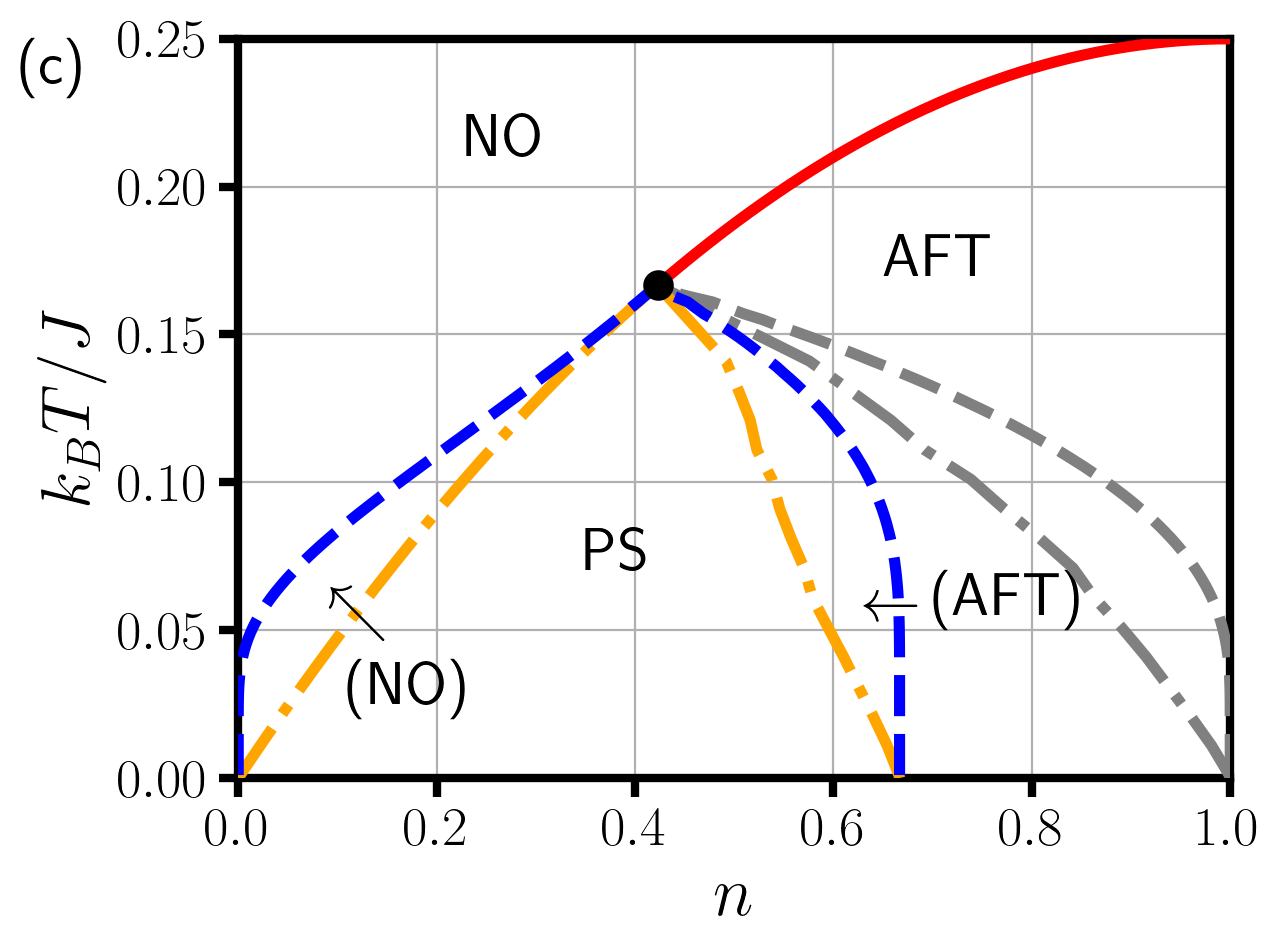}
    \includegraphics[width=\sizethree]{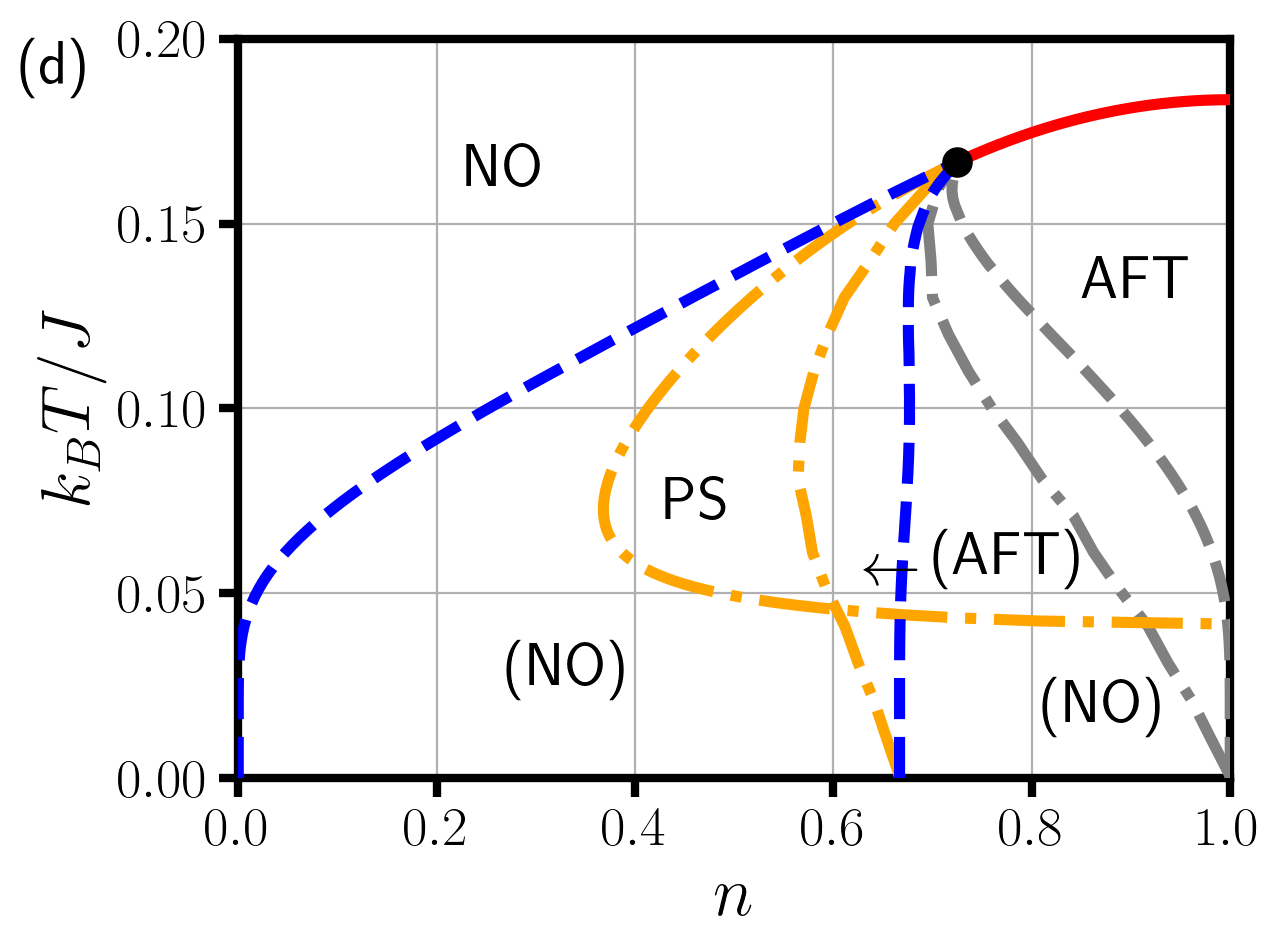}
    \includegraphics[width=\sizethree]{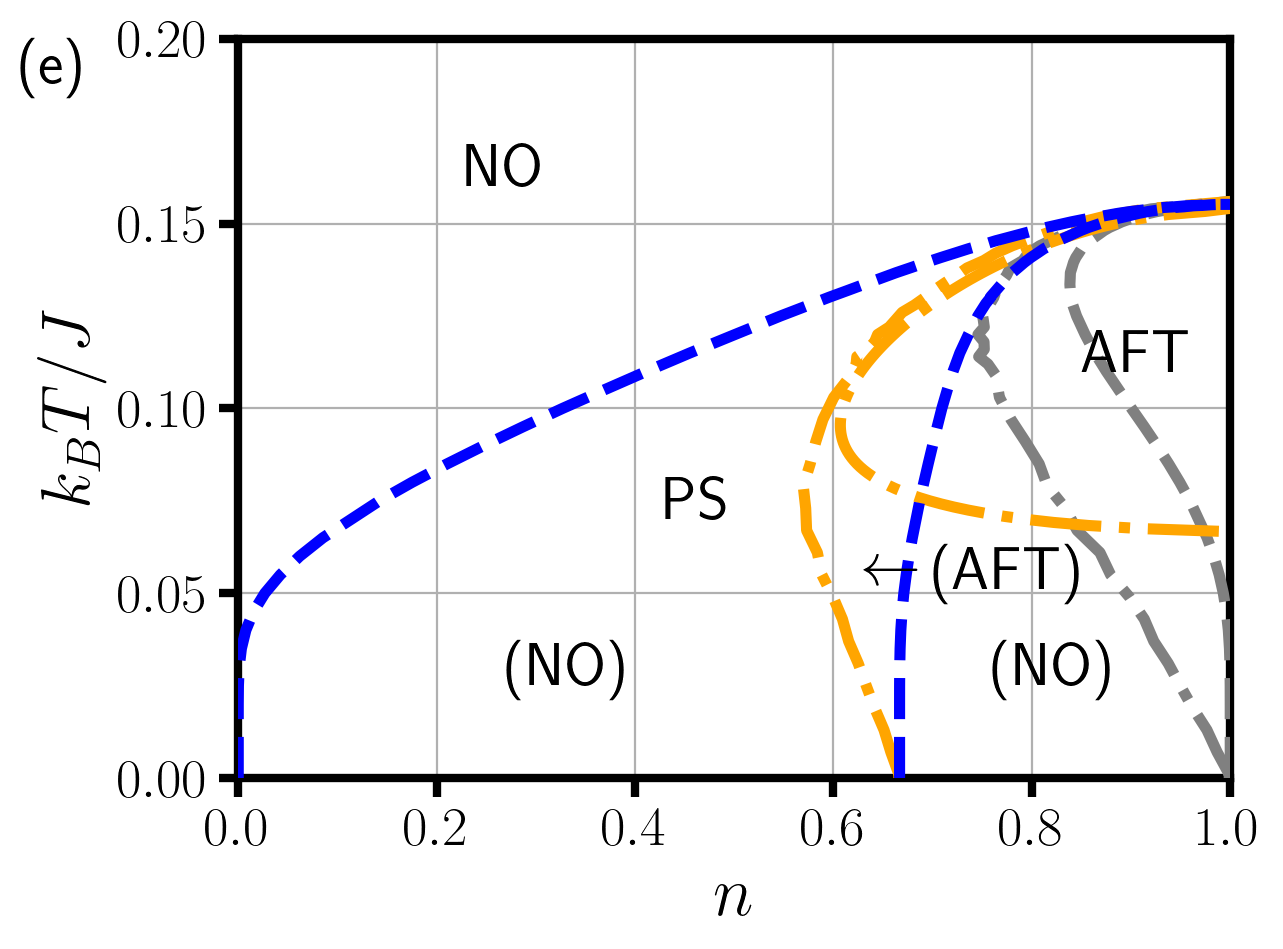}
    \includegraphics[width=\sizethree]{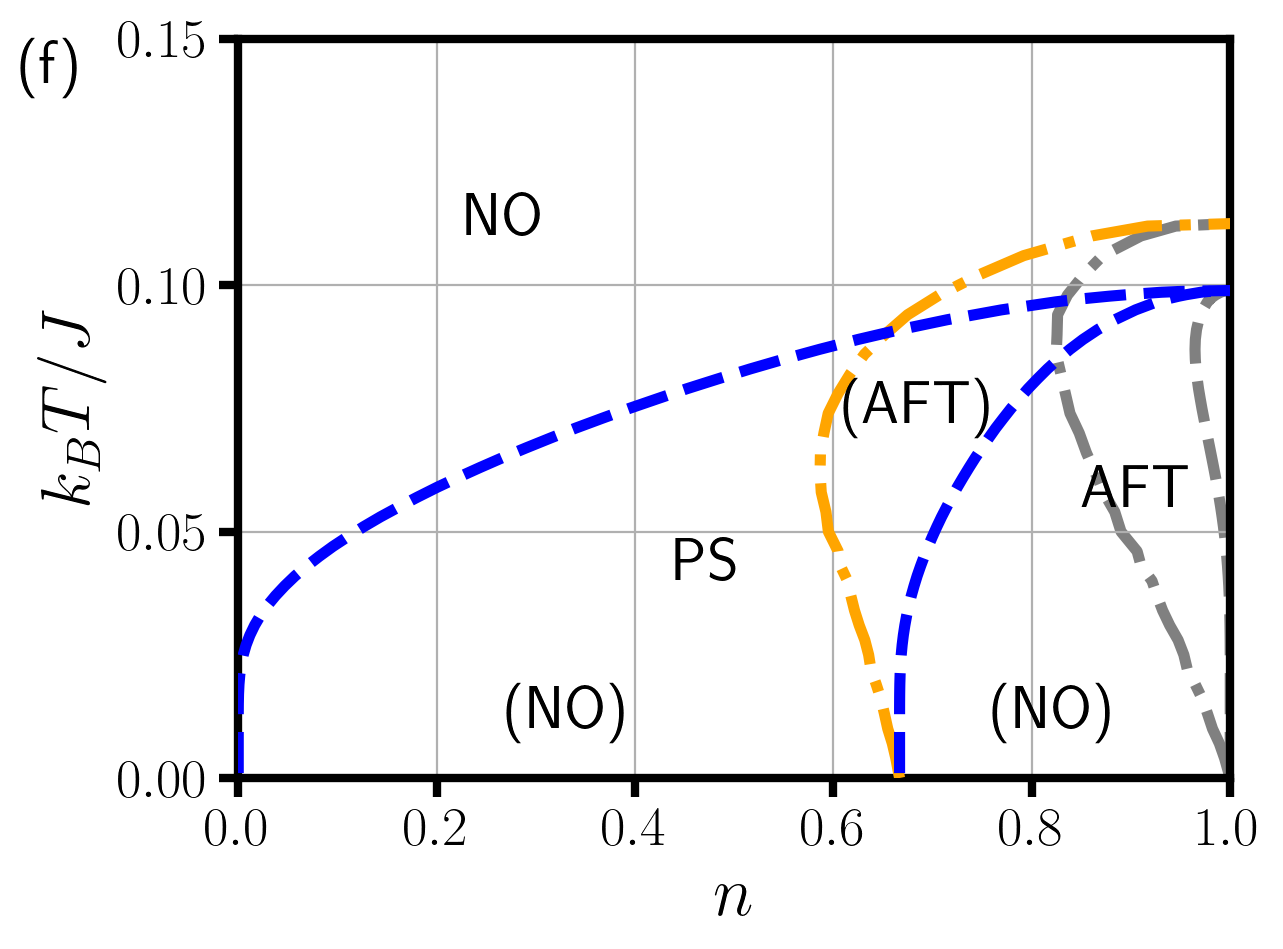}
    \caption{\label{fig:finkTvsn}%
        Phase diagrams for (a) $U/J=5.0$, (b) $U/J=0.5$, (c) $U/J=0.0$, (d) $U/J=-0.2$, (e) $U/J=-0.25$, and (f) $U/J=-0.35$ ($J>0$ AF) as a function of $n$.
        Dashed blue lines denote the boundaries for the PS state occurrence.
        Grey dotted line denotes line shapes for a case of $J<0$ and the {\FER} phase occurrence ($k_BT/|J|$ and $U/|J|$ should be multiplied by factor $2$ according to Eq. (\ref{eq:transFERAFT1})).
        Other denotations as in Fig. \ref{fig:finkTvsmi}.
    }
\end{figure*}

The phase diagrams as a function of $n$ are presented in Fig. \ref{fig:finkTvsn} obtained by comparing the free energies $f=\omega+\mu n$.
For $U>0$ the diagrams are rather  straightforward (cf. Fig. \ref{fig:finkTvsn}(a)-(c)).
For temperatures above  $k_BT=J/6$, the transition between the {\AFT} and the {\NO} homogeneous phases is continuous.
For temperatures below this threshold, the discontinuous NO-{\AFT}  boundary (for fixed $\bar{\mu}$) splits into two lines (dashed lines in Fig. \ref{fig:finkTvsn}).
Between these two (dashed) lines the phase-separated (PS) state occurs.
This state is characterized by the coexistence of two domains (of the {\AFT} and {\NO} phases) in the system (macroscopic phase separation). 
More details about such PS states and application of the Maxwell's construction for correlated system can be found, e.g., in Refs. \cite{ArrigoniPRB1991,bak2004mixed,KapciaPhysA2015,KapciaPRE2017,KapciaNano2021}.
We note that, inside the region of the PS occurrence, there are subregions, where homogeneous {\NO} and {\AFT} phases (on the left and right side, respectively, of the PS region) are metastable (they have higher free energy than that of the PS state). 
For $T=0$, the metastability boundaries coincide with the PS boundaries at $n=0$ (the {\NO} phase metastability) and $n=2/3$ (the {\AFT} phase metastability) and they merge again at the TCP point ($k_BT=J/6$).
Between the dash-dotted lines  homogeneous phases are unstable.

For $U<0$ the situation is much more complex.
Although dash-dotted line marking the boundary of metastability of the {\AFT} phase still coincides with the PS boundary at $T=0$, this is no longer true for the {\NO} metastability line. 
At $U=0$ this boundary at $T=0$ appears at concentration $n=1$. 
For any finite and negative $U$, the {\NO} is metastable at full range of concentrations at low temperatures. 
This applies also to the parameters outside the PS state occurrence region, where the {\AFT} phase is a stable solution. 
This region expands to higher temperatures with decreasing if $U$ (Fig.~\ref{fig:finkTvsn}(d)). 
For $U/J<-1/3 \ln 2 \approx 0.23$, the second order boundary vanishes together with the TCP.
Two PS state boundaries merge at concentration $n=1$ and the {\AFT} metastablity boundary gets its maximum also at $n=1$, but at higher temperatures (Fig. \ref{fig:finkTvsn}(e)).
For $U/J< - 0.279$ the NO metastability boundary vanishes and the NO phase is (meta)stable for any $n$ (Fig. \ref{fig:finkTvsn}(f)).
For $U/J< - 1/2$, the PS state  does not occur any longer, but still, the {\AFT} phase is metastable for $U/J>-1$.
In such conditions, the metastable {\AFT} phase exists in the region of the {\NO} phase stability.

For the case of $J<0$, the boundaries can be obtained from results presented above  by using  simple transformations (\ref{eq:transFERAFT1})-(\ref{eq:transFERAFT2}).
If the results for $J>0$ are qualitatively different from these in Figs. \ref{fig:finkTvsn}, they are marked in the figure by gray dashed and dotted lines. 
In fact, this applies exclusively to the PS-{\FER} boundary and the {\FER} metastability line.
Note also that the mean-field solutions of the investigated model  for $n=1$ can be mapped into the mean-field solutions of other lattice models (various extended Hubbard models in the atomic limit), see e.g., Refs.~\cite{KapciaJSNM2014,KapciaPhysA2015} and references therein. 

\section{Conclusions and final remarks}

In the present work, we investigated phase diagrams of the generalization of the Ising model on the triangular lattice.
Despite the simplicity of model (\ref{eq:ham}) and used approximation, the results found are not trivial.
We determined the phase diagrams for both signs of magnetic interactions at the ground state (rigorous results) and for finite temperatures (the mean-field approximation).
For antiferromagnetic coupling, the ordered phase  (the {\AFT} phase), where the magnetic order (with $s_A = -s_B \neq 0 $ and $s_C=0$) coexists with the charge order  (non-homogeneous distribution of the particle concentration, $n_A = n_B \neq n_C$, for $n\neq 1$), occurs.
This coexistence of these two orders is a consequence of the geometrical frustration of the analyzed system.
In contrary, for ferromagnetic coupling, only magnetic order exists (the {\FER} phase with $s_A=s_B=s_C$ and $n_A=n_B=n_C$), similarly as for a case of bipartite lattices.
For fixed particle concentration, the phase separation state occurs away from the half-filling.
Additionally, the regions for metastable phases were determined.
These regions results from discontinuous transitions existing in the system.

\section*{Acknowledgements}

K.J.K. thanks the Polish National Agency for Academic Exchange for funding in the frame of the Bekker programme (PPN/BEK/2020/1/00184).
K.J.K. is also grateful to prof. Beata Ziaja-Motyka and Center for Free-Electron Laser Science CFEL (Deutsches Elektronen-Synchrotron DESY, Hamburg, Germany)  for hospitality during a part of the work on this project. 

\section*{Declaration of Competing Interest}
The authors declare that they have no known competing financial interests or personal relationships that could have appeared to influence the work reported in this paper.
The funders had no role in the design of the study; in the collection, analyses, or interpretation of data; in the writing of the manuscript, or in the decision to publish the results.

\printcredits

\bibliographystyle{elsarticle-num}
\bibliography{biblio}

\end{document}